\documentclass[doublecol]{epl2}
\usepackage{amsmath}
\usepackage{amssymb}
\usepackage{graphicx}


\title{Heat Transfer between Anisotropic Nanoparticles:\\ Enhancement and Switching}
\shorttitle{Heat Transfer between Anisotropic Nanoparticles}
\date{\today}

\author{Roberta Incardone\inst{1} \and Thorsten Emig\inst{2} \and Matthias Kr\"{u}ger\inst{1}}
\shortauthor{R. Incardone, T. Emig, M. Kr\"uger}

\institute{ 
  \inst{1} Max-Planck-Institut f\"{u}r Intelligente Systeme, Heisenbergstr. 3, D-70569
Stuttgart, Germany and Institut f\"{u}r Theoretische Physik IV, Universit\"{a}t
Stuttgart, Pfaffenwaldring 57, D-70569 Stuttgart, Germany.\\
  \inst{2} Laboratoire de Physique Th\'eorique et Mod\`eles
  Statistiques, CNRS UMR 8626, B\^at.~100, Universit\'e Paris-Sud, 91405
  Orsay cedex, France\\ 
}

\abstract{ We theoretically study heat transfer between two anisotropic nanoparticles in vacuum, and derive closed expressions in terms of the anisotropic dipole polarizabilities.  We show that transfer between two small spheroids can be many times as large as the one for two spheres of same volumes.  Such increase with anisotropy is also found for the heat emission of an isolated small spheroid.  Furthermore, we observe a strong dependence of transfer on the relative orientation, yielding the interpretation as a heat transfer switch. The switch quality, given as the ratio of transfer in the ``on'' and ``off'' positions, is observed to be as large as $10^3$ in the near field and even larger in the far field.
}

\begin{document}
\maketitle

The phenomenon of radiative heat transfer is of amplified interest due to recent experimental observations \cite{Domoto70,Sheng09,
Rousseau09,Ottens11} of its strong increase for distances  below the micron range. In this regime, transfer is enhanced by so called near-field effects attributed to evanescent waves \cite{Polder71}. On the theoretical side, two frameworks underlie thermal radiation and transfer, both of which are fundamental concepts. First, the theory of quantum thermal fluctuations that goes back all the way to the beginning of quantum mechanics, i.e., Planck's law of black body radiation \cite{Planck}. Second, the scattering of light by objects that are small compared to the wavelength, which is by itself a modern field of both experimental and theoretical study \cite{Bohren,Tribelsky2006}.

Radiative energy exchange between objects at different temperatures is on the macro scale well understood in terms of the laws by Planck and Stefan Boltzmann and inclusion of view factors and gray factors \cite{modest} to account for non-planar geometries and non-black bodies, respectively. However, such heat transfer is distinctly different, if the size of the objects or the distance between them is small or comparable to the thermal wavelength, which is roughly 8 microns at room temperature. On these scales, also nontrivial dependencies on the shape of the objects have been observed, as e.g. for sharp tips \cite{McCauley}.  Many recent works computed the exact heat transfer between non planar objects including two spheres \cite{Narayanaswamy08,Sasihithlu11} or a sphere~\cite{Kruger11,Otey,McCauley} or cone \cite{McCauley} in front of a planar surface, periodic structures~\cite{Rodriguez11}, or even more abstract geometries \cite{Rodriguez, RodriguezRe13}. Formalisms for treating fluctuation electrodynamics for arbitrary objects at different temperatures have been recently presented~\cite{Messina,Kruger11,Rodriguez,Narayanaswamy13}.

Due to theoretical simplicity, a large influence on understanding was provided by the study of nano-particles, i.e.,
particles  much smaller than the wavelength \cite{Volokitin01,Chapuis2008,Kruger12}, including many body effects \cite{Messina13}. Such particles are accesibble experimentally. A recent work studies transfer between an anisotropic nano-particle (a spheroid) and a
planar surface \cite{Huth2010} (and the related Casimir interactions between small ellipsiods are analyzed in Ref.~\cite{Emig08}).

In this letter, we study radiative heat transfer between two anisotropic particles, as for example spheroids. We show that
the transfer between two spheroids as well as the heat emission of an isolated one can be many times as large as the
corresponding value for spheres of equal volume. We also demonstrate that the transfer between parallel
spheroids can be many times as large as compared to the perpendicular configuration, an effect which is due to the narrow
peaks of the polarizabilities of nano-particles as a function of frequency. We analyze means of tuning the extent of this
effect by changing the shape or materials of the two objects.

Consider two arbitrary objects at temperatures $T_1$ and $T_2$ in vacuum, whose classical scattering properties in isolation are given in terms of the scattering matrices $\mathcal{T}_{i,\mu\mu^{\prime}}$ \cite{Tsang,Kruger12} (with $i=1,2$ labeling the objects) relating the amplitude of an incoming wave with index $\mu'$ to the scattered  wave with index $\mu$. 
The transfer between the two objects  is then computed as an integral in frequency $\omega$ and the trace of a matrix product \cite{Kruger12},
\begin{align}
\notag H=&\frac{2\hbar}{\pi}\intop_{0}^{\infty}d\omega\left(\frac{\omega}{e^{\frac{\hbar\omega}{k_{B}T_1}}-1}-\frac{\omega}{e^{\frac{\hbar\omega}{k_{B}T_2}}-1}\right)\times\\
&\times\text{Tr}\left[\mathcal{R}_2\mathcal{W}_{21}\mathcal{R}_1^{\dagger}\mathcal{W}_{21}^{\dagger}\right]\label{eq:5-1},
\end{align}
where $\mathcal{R}_i=\frac{\mathcal{T}_{i}+\mathcal{T}_{i}^{\dagger}}{2}+\mathcal{T}_{i}\mathcal{T}_{i}^{\dagger}$ are the emission or absorption matrices. Furthermore, multiple scattering of waves between the
objects is accounted for by the matrix
$\mathcal{W}_{21}=\frac{1}{I-\mathcal{U}_{21}\mathcal{T}_{1}\mathcal{U}_{12}\mathcal{T}_{2}}\mathcal{U}_{21}$, including the
matrix $\mathcal{U}_{21}$. The latter translates the vector waves from the coordinate system of object 1 to the one
of object 2. $\hbar$ and $k_{B}$ are Planck's and Boltzmann's constant, respectively, and the superscript $\dagger$ denotes the adjoined of an operator, e.g. $\mathcal{U}^\dagger_{21,\mu\mu'}=\mathcal{U}^*_{12,\mu'\mu}$.

Eq.~\eqref{eq:5-1} is exact but can often not easily be evaluated analytically. However simplifications are possible
for nano-particles.  If the sizes of the objects, $R_i$ (loosely defined as the largest dimension of the anisotropic
objects), is much smaller than the distance $d$ between their centers, $R_i\ll d$, a one reflection approximation, amounting
to setting $\mathcal{W}_{21}=\mathcal{U}_{21}$ in Eq.~(\ref{eq:5-1}), becomes asymptotically exact.

\begin{figure}
\begin{center}  \includegraphics[width=0.7\linewidth]{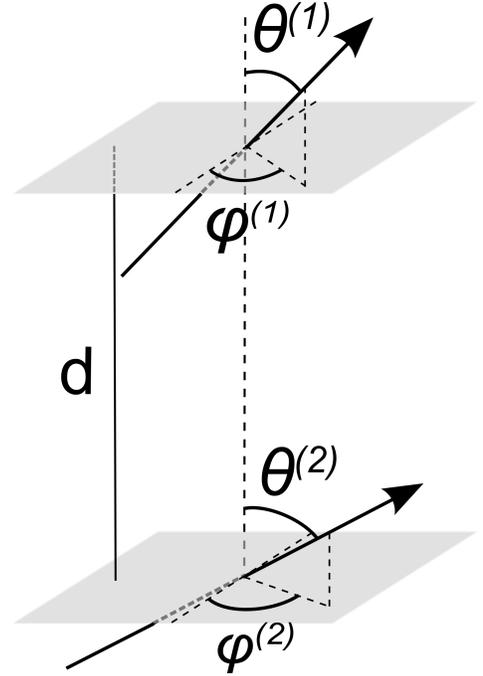}\end{center}
  \caption{\label{fig:1} Sketch of the two objects under study, their center-to-center distance $d$ and orientations ($\theta^{(i)}$ and $\varphi^{(i)}$). The objects are mimicked by arrows, which point in the direction of their respective axes of rotational symmetry, thereby completely specifying the objects' orientations. Note that the resulting transfer will only depend on the difference $\varphi^{(2)}-\varphi^{(1)}$, denoted $\beta$ below. }
\end{figure}

Additional simplification is possible for objects that are sufficiently small such that their scattering properties are
described by their (electric) dipole polarizabilities. While for dielectric objects, this limit is typically fulfilled if $R_i$ is
small compared to the thermal wavelength $\lambda_T=\hbar c/k_BT$, care has to be taken for metallic objects, where the skin
depth is small, and even for $R_i$ of a few nanometer, the description in terms of the dipole polarizabilities is inaccurate
\cite{Kruger12}. Assuming small enough objects, we start by considering the anisotropic dipole polarizability tensors $\hat\alpha$ of the two objects, which, in properly chosen coordinate systems, can be assumed diagonal.
For simplicity, we furthermore restrict to objects with an axis of rotational symmetry, for which the objects polarizabilities are characterized by the two components, 
\begin{align}\label{eq:alpha}
\alpha_\parallel\equiv {\rm Im} [ \hat{\bf e}_\parallel^T\cdot \hat\alpha \cdot\hat{\bf e}_\parallel], \hspace{1cm} \alpha_\perp\equiv {\rm Im} [ \hat{\bf e}_\perp^T\cdot \hat\alpha \cdot\hat{\bf e}_\perp],
\end{align}
where $\hat{\bf e}_\parallel$ and $\hat{\bf e}_\perp$ are unitvectors pointing along the axis of rotational symmetry and perpendicular to it, respectivly.
Our results will depend only on the imaginary part of the polarizability, and the definition in Eq.~\eqref{eq:alpha} allows for a compact notation below. The respective orientations of the axes of symmetry of the two objects are then determined by two angles, with respect to the line connecting their centers, and a line perpendicular to it, see Fig.~\ref{fig:1}. Technically, the objects' polarizabilities $\hat\alpha_g$ measured in the frame shown in Fig.~\ref{fig:1} are obtained by rotations of $\hat\alpha$. 
We then expand Eq.~\eqref{eq:5-1} to lowest order in the size of the objects, where the elements of the $\mathcal{T}$ matrix, measured in the global frame, are straight forwardly obtained from $\hat\alpha_g$, e.g., using spherical vector waves. The dominant elements are for electric polarization and multipole index $l=1$ (dipoles), and are proportional to $(\omega/c)^3$ and linear combinations of the elements of $\hat\alpha_g$. In contrast to a homogeneous sphere, the $\mathcal{T}$ matrix is not diagonal for anisotropic particles.  
Equation~(\ref{eq:heat}) shows the result for the transfer between sufficiently small nano-particles\footnote{As mentioned above, Eq.~\eqref{eq:heat} is valid if ${\cal R}_i$ is small compared to $d$, the material skin depth and the thermal wavelength $\lambda_T$.} (setting $T_2=0$ for brevity), 
\begin{widetext}
\begin{equation}
\begin{split}
\centering
\displaystyle 
&H=\frac{2\hbar}{\pi}\intop_{0}^{\infty}d\omega\frac{\omega^{7}/c^6}{\exp\left(\frac{\hbar\omega}{k_BT_1}\right)-1}\Biggl\{\left(\frac{c^{2}}{\omega^{2}d^{2}}-\frac{c^{4}}{\omega^{4}d^{4}}+\frac{c^{6}}{\omega^{6}d^{6}}\right)
\left(\begin{array}{c}
( \boldsymbol{\alpha}^{(1)} \cdot \textbf{b}_1)(\boldsymbol{\alpha}^{(2)} \cdot \textbf{b}_2)+\alpha_{\perp}^{(1)}\alpha_{\perp}^{(2)}\\
(\boldsymbol{\alpha}^{(1)} \cdot \textbf{b}_1)\alpha_{\perp}^{(2)}+(\boldsymbol{\alpha}^{(2)} \cdot \textbf{b}_2)\alpha_{\perp}^{(1)} 
\end{array}\right) \cdot\left(\begin{array}{c}
\cos\beta\\
\sin\beta
\end{array}\right)\\
&+4\left(\frac{c^{4}}{\omega^{4}d^{4}}+\frac{c^{6}}{\omega^{6}d^{6}}\right) (\boldsymbol{\alpha}^{(2)} \cdot \textbf{b}_2)(\boldsymbol{\alpha}^{(1)} \cdot \textbf{b}_1)-\frac{1}{4}\frac{c^{6}}{\omega^{6}d^{6}}\sin2\theta^{(2)}\sin2\theta^{(1)}(\boldsymbol{\alpha}^{(1)} \cdot \textbf{c}) (\boldsymbol{\alpha}^{(2)} \cdot \textbf{c}) \cos\beta\Biggr\}.\label{eq:heat}
\end{split}
\end{equation}
\end{widetext}
where $\textbf{b}_i=\left(\cos\theta^{(i)},\sin\theta^{(i)}\right)^T$, and we have introduced a vector containing $\alpha_{\parallel}$ and $\alpha_\perp$, $\boldsymbol{\alpha}=\left(\alpha_{\perp},\alpha_{\parallel}\right)^T$. The azymutal angles enter Eq.~\eqref{eq:heat} only through their difference, denoted $\beta=\varphi^{(2)}-\varphi^{(1)}$, and finally $\textbf{c}=\left(1,-1\right)^T$.
Eq.~\eqref{eq:heat}  can be rationalized by comparison to classical radiation
\cite{Jackson}; For example, for all angles being zero, we have the far field ($1/d^2$) term being proportional
to $\sim \alpha_{\perp}^{(1)}\alpha_{\perp}^{(2)}$, a
form which can be anticipated from the radiation field of a dipole \cite{Jackson}. 
A special case of Eq.~\eqref{eq:heat}, evaluated in the figures below, considers the two axes of symmetry in the plane perpendicular to the center-to-center-vector ($\theta^{(i)}=\pi/2$, see sketch in Fig.~\ref{Figure 2}), for which Eq.~\eqref{eq:heat} simplifies,
\begin{equation}\label{eq:fin}
\begin{split}
&H=\int_{0}^{\infty}d\omega\frac{\frac{2\hbar\omega^{7}}{\pi c^6}}{e^{\frac{\hbar\omega}{k_BT_1}}-1}\Biggl\{4\, \alpha_{\perp}^{(2)}\alpha_{\perp}^{(1)}
\left[ \left(\frac{c}{\omega d}\right)^4+\left(\frac{c}{\omega d}\right)^6 \right]+\\
&\left[ \!\left(\frac{c}{\omega d}\right)^{\!\! 2}\! -\!\left(\frac{c}{\omega d}\right)^{\!\! 4}\!+ \!\left(\frac{c}{\omega d}\right)^{\!\! 6}\right]\!
\!\left(\!\!\!\begin{array}{c}
\alpha_{\perp}^{(1)}\alpha_{\perp}^{(2)}+\alpha_{\parallel}^{(1)}\alpha_{\parallel}^{(2)}\\
\alpha_{\perp}^{(2)}\alpha_{\parallel}^{(1)}+\alpha_{\parallel}^{(2)}\alpha_{\perp}^{(1)} 
\end{array}\!\!\!\right) \!\!\cdot\!\!\left(\!\!\!\begin{array}{c}
\cos\beta\\
\sin\beta
\end{array}\!\!\!\right)\!\!\Biggr\}.
\end{split}
\end{equation}
An interesting example scenario is given by spheroids, i.e., ellipsoids with an axis of rotational symmetry. We denote $R_{\parallel}$ ($R_{\perp}$) the radius parallel (perpendicular) to this axis. The corresponding polarizabilities are given by \cite{Landauel,Tsang,Bohren} (recall that $\alpha_{\parallel}$ and $\alpha_\perp$ denote the imaginary part),
\begin{equation}
\alpha_{\parallel/\perp}(\omega)=\text{Im}\left[\frac{1}{3}\frac{R_{\perp}^{2}R_{\parallel}(\varepsilon(\omega)-1)}{(\varepsilon(\omega)-1)n_{\parallel/\perp}(\eta)+1}\right],\label{aa}\\
\end{equation}
with the depolarizing factors $n_{\parallel/\perp}$, 
\begin{align}
n_{\perp}(\eta)&=\frac{1}{2}(1-n_z(\eta)),\\ 
n_{\parallel}(\eta)&=\begin{cases}\frac{1-\eta^2}{2\eta^3}(\log(\frac{1+\eta}{1-\eta})-2\eta)& \; {\rm if}\;  R_{\perp}<R_{\parallel},\\
           \frac{1+\eta^2}{\eta^3}(\eta-2\arctan(\eta))& \; {\rm if}\; R_{\parallel}<R_{\perp}. 
          \end{cases}\label{eq:n}
\end{align}
$\eta$ is the angular eccentricity of the spheroid,  $\eta^{2}=1-\frac{R_{\perp}^{2}}{R_{\parallel}^{2}}$ for a prolate ($R_{\perp}<R_{\parallel}$) and $\eta^{2}=\frac{R_{\parallel}^{2}}{R_{\perp}^{2}}-1$ for an 
oblate spheroid ($R_{\parallel}<R_{\perp}$), and $\varepsilon(\omega)$ is the dielectric permittivity \footnote[2]{
    In the scattering properties of the spheroid, the wavelength $c/\omega$ is compared to geometric scales (e.g. object size) and material 
    scales (e.g. resonance wavelengths). Eqs.~(\ref{aa}-\ref{eq:n}) are valid to lowest order for small $\omega/c$ with respect to geometric scales and to arbitrary order in
    material scales, i.e., $\varepsilon(\omega)$ is not expanded in $\omega$.}. 
The following figures use temperatures $T_1=300$~K and $T_2=0$~K and a typical dielectric material, SiC, with optical properties given by \cite{Spitzer59}
\begin{align}
\varepsilon(\omega)=\varepsilon_{\infty}\frac{\omega^{2}-\omega_{LO}^{2}+i\omega\gamma}{\omega^{2}-\omega_{TO}^{2}+i\omega\gamma}\label{eq:3},
\end{align}
where $\varepsilon_{\infty}=6.7$, and $\omega_{LO}$, $\omega_{TO}$ and $\gamma$ take values of  $0.12$, $0.098$ and $5.88\times10^{-4}$, all in eV. 

\begin{figure}
\includegraphics[scale=0.27]{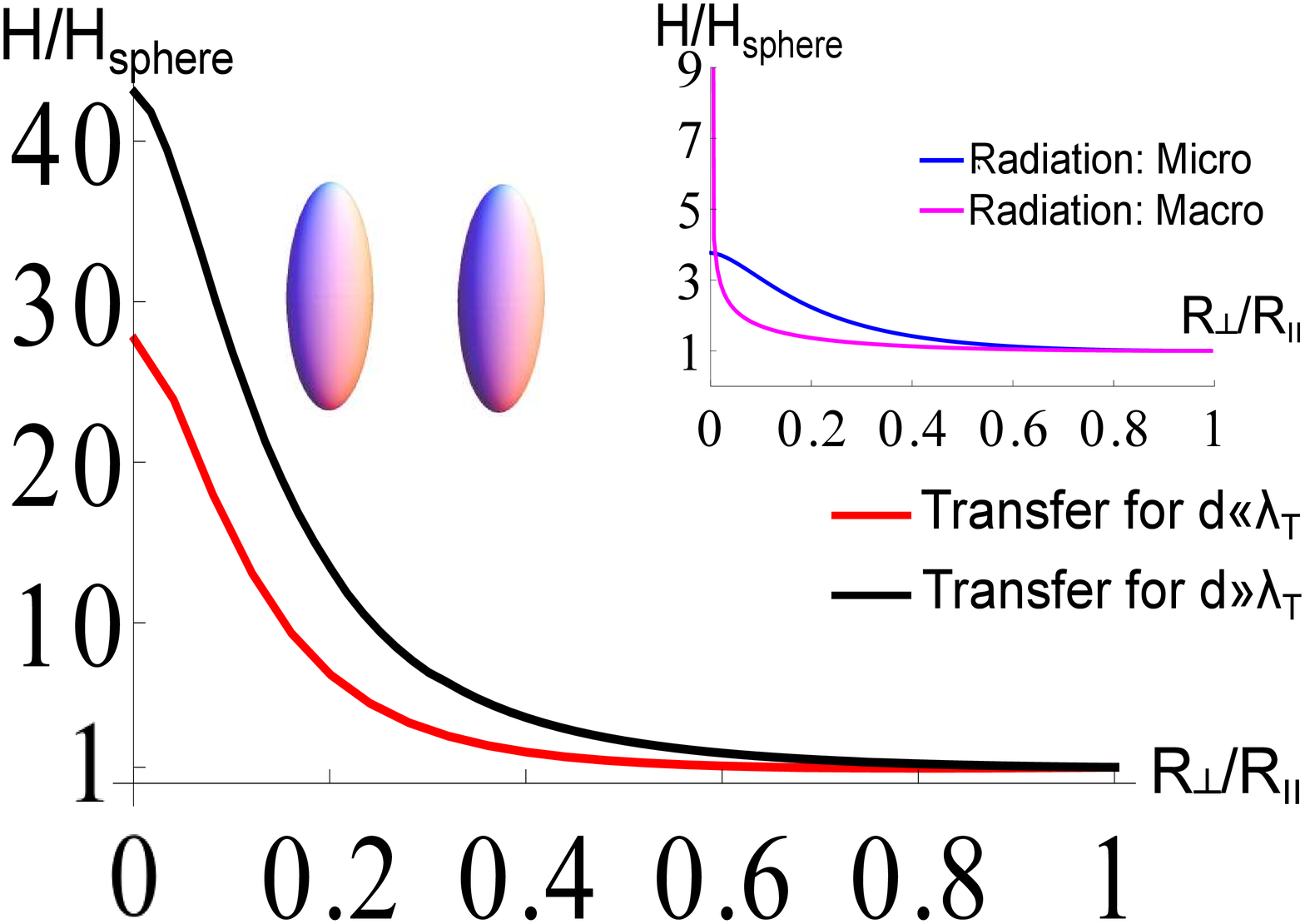}
\caption{Heat transfer between two identical parallel spheroids ($\beta=0$ in Eq.~\eqref{eq:fin}) with  $T_1=300$ and $T_2=0$~K, of fixed volumes  
as a function of $R_{\perp}/R_{\parallel}$ in the limit of small (red curve)  and
large (black curve) distance. In these limits, the curves are independent of distance, see main text. Inset: Dashed blue and magenta lines represent the heat radiation of an isolated micro-spheroid (Eq.~\eqref{eq:rad}) and a macroscopic spheroid (computed by the Stefan-Boltzmann law), respectively. All curves in the figure are normalized by the value for spheres with volumes equal to the spheroid volumes.}\label{Figure 1}
\end{figure}
Figure~\ref{Figure 1} shows the transfer between two {\it identical parallel} spheroids (i.e. $\beta=0$ in Eq.~\eqref{eq:fin}, see also the sketch in the figure) as a function of the ratio $R_{\perp}/R_{\parallel}$ (keeping the volumes fixed), normalized by the value for two spheres of same center-to-center distance and volumes. In other words, the curves approach unity for $R_{\perp}\to R_{\parallel}$, where the polarizabilities in Eq.~\eqref{aa} 
approach the polarizability of a sphere,
\begin{equation}
\underset{\eta\rightarrow0}{\lim}\alpha_{\perp/\parallel}=R_{\perp}^{2}R_{\parallel}\frac{(\varepsilon-1)}{(\varepsilon+2)}.
\end{equation}
The figure shows both the far field regime, $d\gg\lambda_T$, where the $1/d^2$-term in Eq.~\eqref{eq:fin} dominates, as well as the near field regime, where the $1/d^6$-term is largest. In these limits, the distance dependence cancels in the shown ratio, since these limiting power laws are the same for any shape. We note that the transfer in both regimes can be many times (30-40) as large as the transfer between two spheres, showing the strong tunability of transfer by changing the objects' shapes.
 
The inset presents the heat emission of an isolated spheroid which can be either micros- or macroscopic. 
The latter case describes the situation where the spheroid is large compared to the thermal wavelength $\lambda_T$.
For the {\it microscopic} case the emission is given in terms of the trace of the polarizability,
\begin{equation}\label{eq:rad}
H=\frac{2\hbar}{\pi}\int_{0}^{\infty}d\omega\frac{\omega^{4}}{c^{3}(\exp(\frac{\hbar\omega}{kT})-1)}\frac{2}{3}\text{Im}\text{Tr} [\hat\alpha],
\end{equation}
which, due to the cyclic property of the trace, is independent of orientation.
It also increases with decreasing $R_{\perp}/R_{\parallel}$, yet not as strongly as expected from the transfer curves; Naively, we expect the transfer to be the product of the emissivities, i.e., the blue curve in the inset to be roughly the square root of the red or black curves in the main graph. This estimate, which neglects orientation effects, appears however too rough. We labelled the result of Eq.~\eqref{eq:rad} by ``Micro'' as it holds for small anisotropic particles (e.g. spheroids). In contrast, the second curve in the inset of Fig.~\ref{Figure 1} gives the emissivity of a {\it macroscopic} spheroid as a function of its eccentricity. Macroscopic bodies emit proportional to their surface area \cite{modest} according to the Stefan-Boltzmann law, almost independent of shape. The curve labelled ``Macro'' thus shows the surface area of the spheroid for fixed volume. It is below the curve for the micro-spheroid, however diverges for $R_{\perp}/R_{\parallel}\to0$ as $(R_{\perp}/R_{\parallel})^{-\frac{1}{3}}$.   

The inset of Figure~\ref{Figure 2} shows the transfer for two identical spheroids, as a function of  their relative angle $\beta$ (see Eq.~\eqref{eq:fin} and the aketch in the Figure). The curves, drawn for moderately stretched objects with $R_{\perp}/R_{\parallel}=0.2$, show the characteristics of a switch, the transfer in the parallel (``on'') position being more than thousand times as large as in the perpendicular (``off'') position. The main figure shows the switch quality, i.e., the ratio of maximal and minimal values of the transfer in the inset (i.e., the value for $\beta=0$), as a function of eccentricity. Making the spheroids thinner increases the quality, which can, for very stretched objects, reach values of ten thousands. The strong dependence on $\beta$ shown in the inset can be understood by noticing that $\alpha_{\perp}$ and $\alpha_{\parallel}$ show distinct resonances at different frequencies (compare also the inset of Fig.~\ref{fig:2d} below). Regarding $d\gg\lambda_T$, the integrand is $\sim\alpha_{\perp}^{(1)}\alpha_{\perp}^{(2)}+\alpha_{\parallel}^{(1)}\alpha_{\parallel}^{(2)}$ in the ``on'' position, while it is  $\sim\alpha_{\perp}^{(1)}\alpha_{\parallel}^{(2)}+\alpha_{\parallel}^{(1)}\alpha_{\perp}^{(2)}$ in the ``off'' position. The overlap of $\alpha_{\perp}$ and $\alpha_{\parallel}$ is small, reducing the relative transfer in the ``off'' position. The shown behavior thus follows from a correlation of shape and material related properties. For $d\ll\lambda_T$, Eq.~\eqref{eq:fin} shows a  $\beta$-independent term that causes the switch quality to be lower by almost a  factor of a hundred.

\begin{figure}
\includegraphics[scale=0.27]{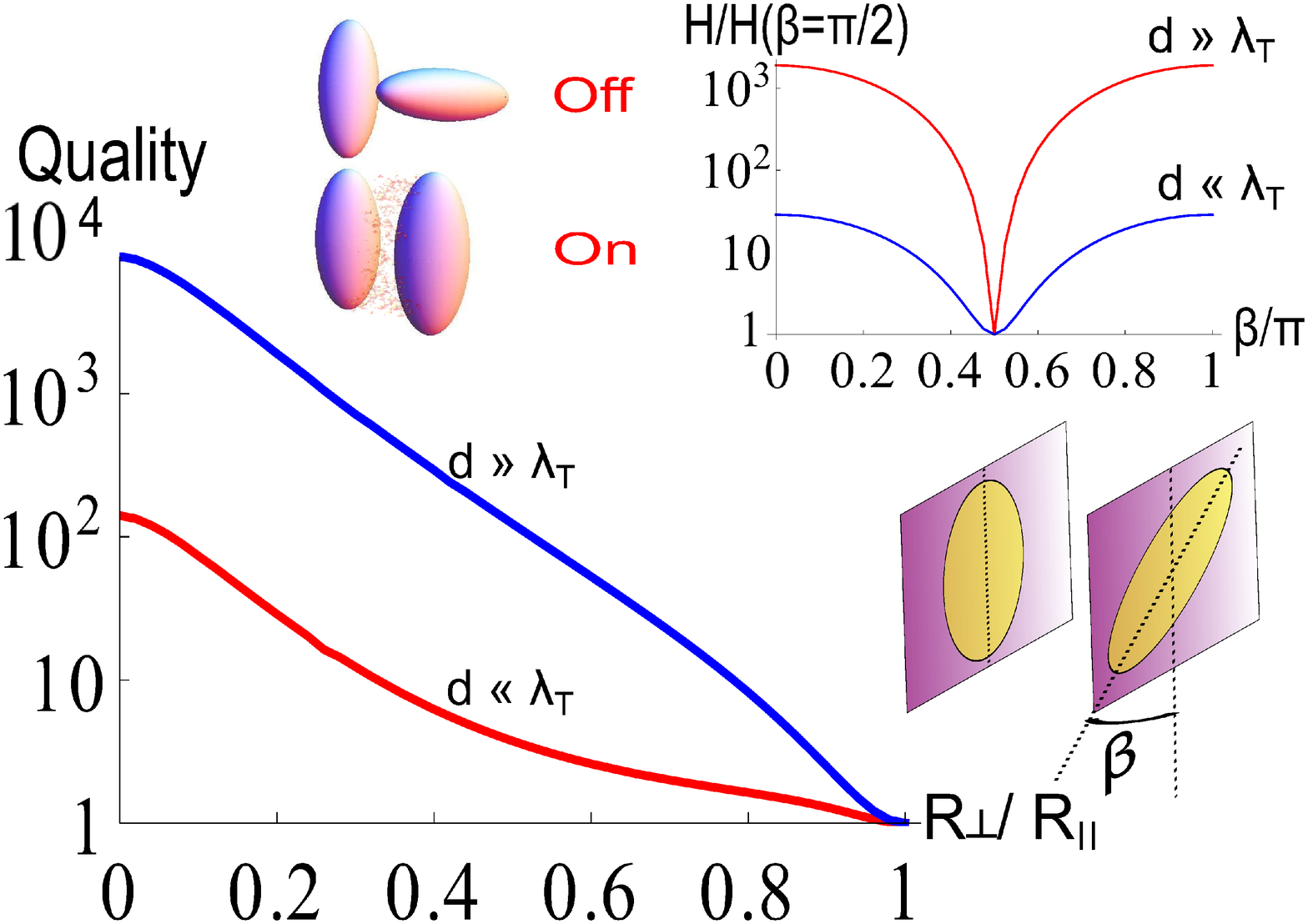}
\centering{}
\caption{Heat transfer between two identical spheroids with $R_{\perp}/R_{\parallel}=0.2$, as a function of angle $\beta$ (see inset). Main graph gives the quality of the transfer switch as a function of $\frac{R_{\perp}}{R_{\parallel}}$ (see main text). In the shown limits for $d$, the transfer assumes simple power laws, such that the given ratios are $d$-independent. }\label{Figure 2}
\end{figure}

This angle independent term can be {\it suppressed} by e.g. using two spheroids that have distinct shape and/or material composition. Figure \ref{fig:2d} shows the case of two spheroids with slightly detuned material resonances of $\sqrt{1.05}\,\omega_{LO}$ and  $\sqrt{1.1}\,\omega_{LO}$, respectively. As seen in the upper inset, this reduces the overlap of the polarizabilities $\alpha^{(1)}$ and $\alpha^{(2)}$, thereby suppressing the angle independent term in Eq.~\eqref{eq:fin}. Furthermore we also chose the anisotropy of the two spheroids slightly different, with $R_{\perp}/R_{\parallel}=0.25$ and $0.2$. This restores the desired overlap of $\alpha_{\parallel}^{(1)}$ with $\alpha_\parallel^{(2)}$ (see inset). These manipulations lead to a switch quality of $\sim 1400$ in the near field, for moderately stretched objects.

\begin{figure}
\includegraphics[scale=0.27]{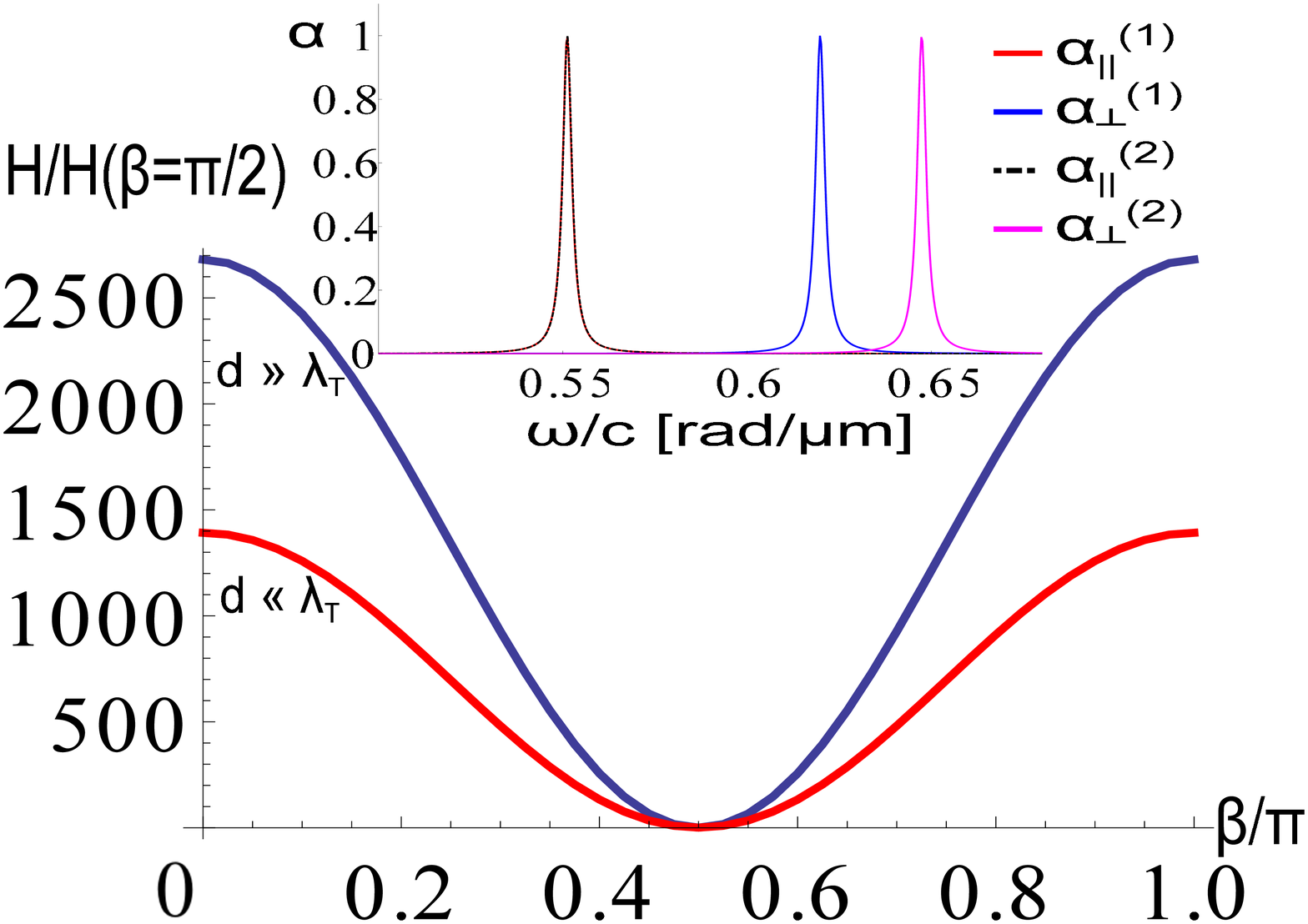}
\centering{}
\caption{Heat transfer between slightly different spheroids, i.e.,  $R_{\perp}/R_{\parallel}=0.25$ (0.2) and $\omega_{LO}^2$ in Eq.~\eqref{eq:3} multiplied by 1.05 (1.1) for object 1 (object 2), as a function of angle $\beta$. Insert: Imaginary part of the polarizabilities (normalized to unity), demonstrating the desired overlaps.}\label{fig:2d}
\end{figure}
\begin{figure}
\includegraphics[scale=0.27]{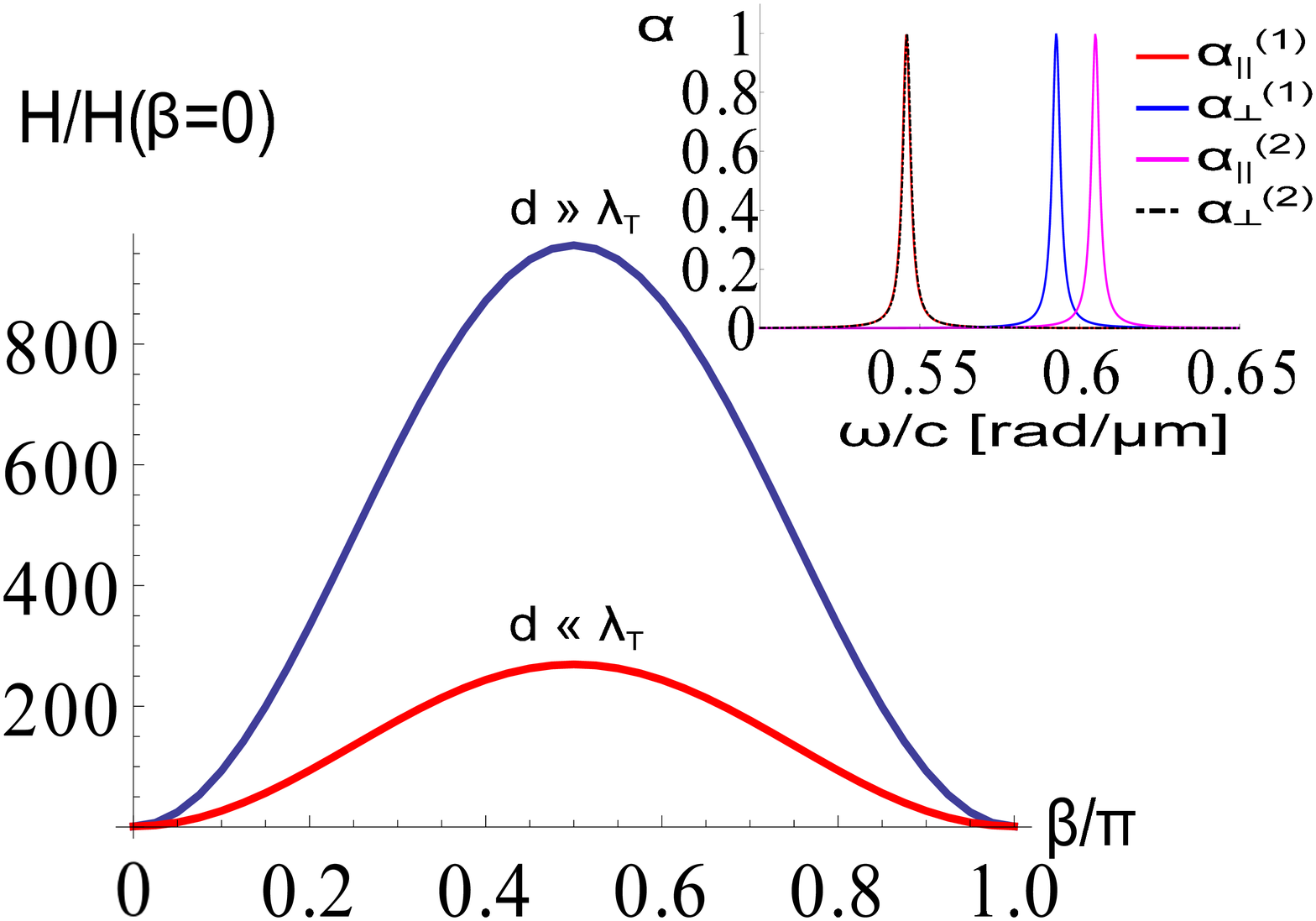}
\centering{}
\caption{Heat transfer between a prolate ($R_{\perp}/R_{\parallel}=0.30$) and an oblate ($R_{\parallel}/R_{\perp}=0.145$) spheroid with identical dielectric permittivities. Insert shows the imaginary part of the polarizability of the two objects (normailzed to unity), demonstrating the desired overlaps.}\label{Figure 3}
\end{figure}

The angle independent term can as well be suppressed by considering transfer between a prolate and an oblate spheroid, as shown in Fig.~\ref{Figure 3}. The inset shows again the different polarizabilities, where we note that the desired overlaps have been achieved for {\it identical} $\varepsilon(\omega)$, i.e., the two spheroids have identical materials. The main panel demonstrates  the strong orientation dependence, where now the transfer is maximal for $\beta=\pi/2$. The quality is $\sim 300$ in the near field, which is still an unexpectedly large value, given that the prolate spheroid is not very much stretched ($R_{\perp}/R_{\parallel}=0.3$).     

Heat transfer between anisotropic nanoparticles allows for a large freedom of tuneability. For a typical dielectric material, the transfer between parallel spheroids  can be $\sim 30-40$ times as large as for two spheres of equal volumes and distance. The strong dependence on the relative angle of two spheroids, which can also be tuned, can increase or reduce the transfer by factors up to $10^3$ or $10^4$ by a simple twist of one of the objects, thereby providing the interpretation of a transfer switch. 

Future work can investigate these effects for metallic particles. Indeed, intriguing scaling laws, slightly different from those found for the equilibrium Casimir force \cite{Emig08}, emerge for sufficiently large $\varepsilon(\omega)$ and small $R_\perp/R_\|$. There exist a regime where from Eq.~\eqref{aa}, $\alpha_\| \sim R_\|^5/(\log^2(R_\|/R_\perp) R_\perp^2)$, and hence the quantity shown in Fig.~\eqref{Figure 1} scales as $H/H_{\rm sphere} \sim (R_\|/R_\perp)^8 / \log^4(R_\|/R_\perp)$ hence strongly increasing for decreasing $R_\perp/R_\|$. At even smaller $R_\perp/R_\|$, this divergence is however cut off and saturates (depending on $\varepsilon(\omega)$).

We thank G.~Bimonte, R.~L.~Jaffe, M.~Kardar, A.~W.~Rodriguez and M.~T.~H. Reid for useful discussions. This research was supported by DFG grant No. KR 3844/2-1.

\end{document}